\documentclass[conference]{IEEEtran}
\IEEEoverridecommandlockouts

\usepackage{cite}
\usepackage{amsmath,amssymb,amsfonts}
\usepackage{algorithmic}
\usepackage{graphicx}
\usepackage{textcomp}
\usepackage{xcolor}
\def\BibTeX{{\rm B\kern-.05em{\sc i\kern-.025em b}\kern-.08em
    T\kern-.1667em\lower.7ex\hbox{E}\kern-.125emX}}

\newtheorem{definition}{Definition}[section]

\begin{document}

\title{Secret Sharing in 5G-MEC: Applicability for joint Security and Dependability\\

\thanks{This work was supported by the Research Council of Norway through the 5G-MODaNeI project (no. 308909).}
}

\author{\IEEEauthorblockN{1\textsuperscript{st} Thilina Pathirana}
\IEEEauthorblockA{\textit{Department of Electrical Engineering and Computer Science} \\
\textit{University of Stavanger)}\\
Stavanger, Norway \\
thilina.pathirana@uis.no}
\and
\IEEEauthorblockN{2\textsuperscript{nd} Ruxandra F. Olimid}
\IEEEauthorblockA{\textit{Department of Computer Science} \\
\textit{University of Bucharest)}\\
Bucharest, Romania \\
ruxandra.olimid@fmi.unibuc.ro}
}

\maketitle

\begin{abstract}
Multi-access Edge Computing (MEC), an enhancement of 5G, processes data closer to its generation point, reducing latency and network load. However, the distributed and edge-based nature of 5G-MEC presents privacy and security challenges, including data exposure risks. 
Ensuring efficient manipulation and security of sensitive data at the edge is crucial. To address these challenges, we investigate the usage of threshold secret sharing in 5G-MEC storage, an approach that enhances both security and dependability. A $(k,n)$ threshold secret sharing scheme splits and stores sensitive data among $n$ nodes, requiring at least $k$ nodes for reconstruction. The solution ensures confidentiality by protecting data against fewer than $k$ colluding nodes and enhances availability by tolerating up to $n-k$ failing nodes. This approach mitigates threats such as unauthorized access and node failures, whether accidental or intentional. We further discuss a method for selecting the convenient MEHs to store the shares, considering the MEHs' trustworthiness level as a main criterion. Although we define our proposal in the context of secret-shared data storage, it can be seen as an independent, standalone selection process for 5G-MEC trustworthy node selection in other scenarios too.
\end{abstract}

\begin{IEEEkeywords}
5G-MEC, security, dependability, secret sharing
\end{IEEEkeywords}

\section{Introduction}

\label{sec:introduction}

The fifth-generation mobile network (5G) technology allows high speed, low latency, and reliable data transfers. This facilitates scenarios like the Internet of Things (IoT), smart cities, self-driving cars, and other applications. Multi-access Edge Computing (MEC), previously known as Mobile Edge Computing \cite{mec2014,ETSI:5G:MEC}, enriches the 5G’s potential as the computations are performed at the edge. By design, MEC minimizes overall latency and makes optimum use of the bandwidth resources by performing work close to the source. Similarly, local storage is a real benefit present in several IoT applications, including smart homes, wearables, healthcare, environment monitoring, and farming \cite{liyanage2021driving,chattopadhyay2024secret}.
As a consequence, MEC has lowered its reliance on centralized cloud architectures. 

On the other side, because of the distributed and decentralized nature of the 5G-MEC environment, as well as its placement in insecure physical locations and the use of wireless connectivity, data storage at the edge is vulnerable to various security threats. The edge nodes are often placed in unguarded or even hard-to-secure locations so that they can face physical attacks due to easy access. Attacks such as Distributed Denial of Service (DDoS) can flood the infrastructure by attacking several edge nodes and thus disrupting the service and data availability. Furthermore, edge nodes are equipped with limited capabilities that restrict the adopted security protocols, which, again, puts them at risk. All this becomes even more significant as the end-users normally own the data collected at the edge, so data is highly sensitive. This is the case of healthcare, financial, or any other personal data that is processed at the edge. Data protection laws (for instance, GDPR \cite{gdpr}) pose a challenge in a distributed edge environment since there is a need to enforce rules on how data is stored and processed.

In this context, over the years, different cryptographic methods have been investigated and used. In particular, secret sharing \cite{chattopadhyay2024secret} has proven its applicability in many scenarios, including networking.
A Secret Sharing Scheme (SSS) splits into several shares and further reconstructs the original data at need, using a qualified set of shares. Shamir's (as any threshold SSS) asks for any set qualified to reconstruct to have at least a given threshold $k$ of shares~\cite{shamir1979share}. 5G-MEC can employ its decentralized design to store these shares over the edge nodes. By construction, the solution facilitates theoretical perfect data secrecy when fewer than $k$ nodes are compromised and failure tolerance up to $n-k$ failing nodes. 

\subsection{Motivation}
The motivation of this paper is two-fold. 

Firstly, there is a need for 5G-MEC solutions that are jointly secure and dependable. This holds especially in the rise of mission-critical
scenarios such as emergencies (e.g., remote medical emergency, emergency communications) that have strict requirements on both security and dependability: the communication must not be tampered with and must remain functional regardless of intentional attacks or unintentional faults. Hence, looking at building blocks that offer both security and dependability by construction, such as threshold cryptography - in particular, secret sharing - is a natural direction to investigate. Several research questions remain open. Are such primitives appropriate to use in 5G-MEC to offer joint security and dependability by construction? If so, under what scenarios and how can/should they be integrated to maximize success?

Secondly, 5G-MEC data storage brings lower latency and bandwidth consumption than other established solutions, e.g., the cloud. However, this comes at the cost of placing nodes close to end users, which by design increases security risks and decreases storage capabilities. Individual MEC nodes are exposed to intentional attacks and prone to unintentional failures, which can cause sensitive data exposure, data loss and/or unavailability of services. Such an undesired behavior becomes relevant, especially in scenarios where high-volume data are generated continuously, such as smart cities with autonomous vehicles and real-time surveillance. 

\subsection{Novelty and Contributions}

Motivated by the reasons mentioned above, we investigate the utility of secret sharing in 5G-MEC as a cryptographic primitive that provides both data security and redundancy by construction. We then focus on 5G-MEC data storage and investigate to what extent secret sharing, in particular, Shamir's scheme \cite{shamir1979share} - could help satisfy both security and dependability needs. We investigate an algorithm to select the 5G-MEC nodes used in the sharing to maximize trustworthiness and availability and decrease communication latency. We perform a theoretical analysis and conduct a basic testbed experiment using Simu5G~\cite{nardini2020simu5g}. 
Our main contributions are as follows.
\begin{itemize}
    \item We investigate how secret sharing was used in the literature to enable security and/or dependability in 5G-MEC. 
    \item We integrate secret sharing in the context of 5G-MEC data storage as a solution that brings data security and redundancy by construction and discuss its feasibility in terms of security, dependability, and performance. Our analysis encompasses theoretical and practical methods.
    \item We discuss a method for the 5G-MEC node selection, considering the nodes' trustworthiness level and their capabilities. Our proposal is defined in the context of secret-shared data storage. However, in other scenarios, it can also be seen as an independent, standalone selection process for 5G-MEC trustworthy node selection.  
\end{itemize}

\subsection{Outline}

The paper is organized as follows. Section~\ref{sec_relw} briefly discusses the existing literature in two directions: (1) secret sharing utility in 5G-MEC and (2) storage solutions, which presents (A) distributed storage in 5G-MEC and (B) the role of secret sharing in distributed storage. Section~\ref{sec_background} gives the preliminaries in terms of 5G-MEC storage solutions and secret sharing. Section~\ref{sec_solution} proposes a secure storage solution for MEC based on Shamir's scheme, including an implementation exemplification and discussing a process for selecting the MEHs to store the shares. Section~\ref{sec_eval} discusses the proposed solution in terms of security and dependability, with a focus on performance and introduced complexity. Section~\ref{sec_conclusion} concludes.

\section{Related work}
\label{sec_relw}

\subsection{Secret Sharing in 5G-MEC}

\begin{figure}
    \centerline{\includegraphics[width=.4\textwidth]{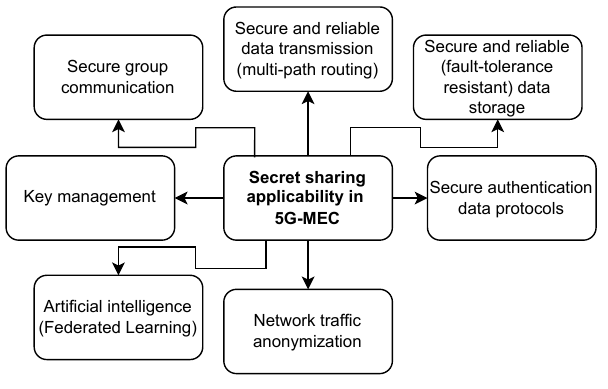}}
    \caption{Applicability of secret sharing in 5G-MEC}
    \label{fig:applicability}
\end{figure}

In the literature, secret sharing was proposed to improve several aspects concerning privacy and security in 5G-MEC. Fig.\ref{fig:applicability} illustrates the applicability of secret sharing in 5G-MEC, even though some aspects are not explicitly discussed in the literature in the exact context of 5G-MEC. We thus focus on 5G-MEC but sometimes refer to related scenarios (general MEC or 6G) that can be, to some extent, applicable to 5G-MEC. Note that the existing literature on secret sharing usage in networking in general, also in relation to security and dependability, is much larger. In particular, \cite{chattopadhyay2024secret} lists the applicability of secret sharing in close-related domains, including IoT, cloud, and smart grids.

\textit{Data transmission}. 
Liyanage et al. \cite{liyanage20185g} illustrate the use case of the SDN controller that chooses multiple paths in the network to transmit different parts of the data stream. 

Zhao et al. \cite{zhao2024cmt} mention threshold secret sharing as a representative for multipath routing in 6G edge networks, too, as a solution to provide both increased security (data is difficult to be stolen) and reliability (data has a certain level of fault tolerance).

\textit{Network traffic anonymization}. Niewolski et al. \cite{niewolski2023security} propose to anonymize six network parameters - the source and destination MAC addresses, IP addresses, and ports - by bitwise xor-ing with shares of a secret, using rules configured on the UPF core and UPF MEC. 

\textit{Key management}. Wang \cite{wang2018privacy} uses Shamir's secret sharing to design a key exchange protocol for secure communications between end devices and edge devices. 

Zhang et al. \cite{zhang2021secure} propose a data storage and sharing scheme for blockchain-based mobile-edge computing. In particular, they use secret sharing to share a private signing key and store the encrypted shares in the blockchain. The solution is debatable because storing sensitive data in the blockchain (in particular private keys) is a bad practice, even if the stored data is encrypted. 
Sun et al. \cite{sun2022fine} propose a data‐sharing model for intelligent terminals, which introduces a key self‐certification algorithm and uses Shamir's secret sharing to secure a key distribution process. The proposed solution uses expensive primitives and technologies such as public key encryption, bilinearity, blockchain, etc. It is worth noting that the advantage of sharing an encryption key (rather than the data) is clear in terms of storage savings. However, this weakens security, making data reveal computationally hard (not theoretically secure, as in the case of direct sharing of the data). In this work, we are not interested in sharing keys but only the data directly.

\textit{Artificial Intelligence}. Ari et al. \cite{ari2024ensuring} mention secret sharing in the context of federated learning for IoT data monitoring and analysis at the edge to allow clients to hide their model contributions to the server. Similarly, Wang et al.\cite{wang2022privacy} design a lightweight privacy protection protocol based on threshold secret sharing and weight masks in the context of federated learning under edge computing for healthcare.

\subsection{Storage Solutions}

We further consider existing storage solutions from two different perspectives: (1) storage in 5G-MEC (sometimes MEC in general) and (2) secret sharing-based storage solutions, regardless of the context.

\subsubsection{Storage Solutions in 5G-MEC}

Al Ridhawi at el. \cite{al2018collaborative} propose a solution that replicates files and service tasks from the cloud to MEC servers. Replication at the MEC servers can be expensive, and thus performed for highly requested data. Moreover, replication is also present at the end user by caching the highly-accessed data. This approach ensures fast data access and creates an interactive environment in 5G due to the possibility of user-MEC collaboration to share content.

Makris et al. \cite{app12178923} present challenges related to distributed storage in the context of edge computing in terms of latency, data accessibility, data accuracy, scalability, interoperability between hardware and software, and security. They point out that storage approaches used in edge nodes require less complexity than those used in cloud nodes ~\cite{app12178923}. They evaluate MinIO~\cite{minio}, BigchainDB~\cite{mcconaghy2016bigchaindb}, and IPFS~\cite{benet2014ipfs}. Each of the mentioned solutions provides a different type of storage (object, blockchain, and file storage), which can be considered suitable for edge computing. MinIO, however, stands out for its decentralization and scalability. While blockchains offer data integrity and transparency, they bring high computational and storage requirements that are not suitable for edge nodes. The study also notes that although encryption improves security, it puts additional pressure on the system resources, which asserts that there is a compromise between efficiency and security in existing edge storage solutions.

According to both ~\cite{makris2022towards} and ~\cite{psomakelis2023lightweight}, distributed storage frameworks that utilize Kubernetes~\cite{baier2017getting} and MinIO~\cite{minio} are proposed to achieve optimal data management in 5G-MEC settings. Although these systems exhibit their efficiency in data processing and durability, they also indicate security constraints, particularly related to encryption and authentication for mobility and handovers in 5G networks. The decentralized storage of data and the implementation of dynamic mechanisms engender both exposure risks and challenges relating to stringent access control, primarily when seeking to protect privacy at 5G-MEC edge nodes. In addition to attempts to enhance security through lightweight solutions, these measures may still prove insufficient for managing the high-rate, low-latency demands of 5G, as they lack broad protection protocols and suitable responses to time-sensitive security risks, particularly when sensitive data is managed across both the cloud and the edge nodes.
In relation to secret sharing, the R-Drive encryption used in \cite{makris2022towards} makes use of 256-bit AES encryption where the key is split using Shamir's SSS (in the all-or-nothing settings). Without arguing here the selection of the Shamir SSS instead of a more efficient all-or-nothing scheme (e.g., XOR-based), we note that this solution fundamentally differs from ours in that it shares the key, not the data itself. Although this approach has its own advantages (e.g., lower processing), the confidentiality remains computational.

Sagor et al.~\cite{sagor2024distressnet} propose a MEC-based storage solution focused on disaster response and tactical scenarios in Cyber-Physical Systems while maintaining the data secure and accessible. Even though the system is effective for mission-critical applications that call for minimal latency and high fault tolerance, it has particular limitations. Using adaptive erasure coding at the edge introduces computational burdens, and depending solely on one master node can lead to data loss or high latency if a failure occurs during important missions.

Chen et al.~\cite{chen2023secure} look into a cloud-edge collaborative fault-tolerant storage method to strengthen fault tolerance and storage efficiency. Their solution depends heavily on the cloud, causing latency issues in 5G-MEC time-sensitive applications. Cloud-based parity storage makes data vulnerable to interception during the transfer.
Adding erasure coding and Software-Defined Networking (SDN) increases complexity and the overhead that reduces performance levels during heavy traffic and critical, real-time 5G-MEC applications.

\subsubsection{Secret Sharing-Based Storage Solutions}

Over the years, secret sharing has been successfully used to secure data storage. Compared to traditional solutions, secret-sharing-based solutions have some considerable advantages. By design, they bring theoretical security\footnote{Up to compromising enough nodes, in case of threshold secret sharing.}, in opposition to the computational secrecy obtained by encryption. They also introduce dependability by default (as the data is split over several nodes and can be recovered even if some nodes are unavailable), thus eliminating the need for additional backup solutions.

Examples include PASIS~\cite{W:2000}, a threshold-secret sharing-based storage system, and GridSharing~\cite{SB:2005}, an all-or-nothing secret sharing-based solution, POTSHARDS~\cite{SGMV:2009}, which combines the principles of the previous two, the Redundant Array of Independent Disks (RAID) distributed algorithms for data recovery,

OceaneStore~\cite{RWE:2001}, used as the storage layer for the ePOST serverless email system \cite{ePost:2014}, Glacier~\cite{HMD:2005}, and  AONT-RS~\cite{RP:2011}, implemented in Cleversafe solutions, later acquired by IBM. Other solutions that use secret sharing for cloud storage (e.g., \cite{bessani2013depsky,framner2019making}) or in blockchain (e.g., \cite{raman2018distributed,sun2022fine}) are also available. 

The literature on storage solutions that use secret-sharing is quite vast, so it is out of our scope here to provide a comprehensive image of the existing works. We further restrict the presentation of the work within the settings of the edge.

Pu et al. \cite{pu2020r2peds} propose a distributed edge storage scheme named {{$R^2PEDS$}}, which makes use of secret sharing. The authors claim the solution is to preserve privacy while allowing for data recovery. However, the proposed solution tries to minimize storage by using secret sharing in a deterministic way defined by the data. An in-depth security evaluation of the proposal is out of our scope. However, this approach has already been proven insecure in the literature \cite{olimid2016security} and should not be followed. To maintain the security properties of Shamir's scheme, the coefficients of the polynomial - except the free term, which equals the shared secret - must be randomly chosen. Hence, the dimension of each share equals the dimension of the shared secret. A direct consequence of this is an overhead in storage. For these reasons, we are not interested in comparison with \cite{pu2020r2peds}.


\section{Preliminaries}
\label{sec_background}


\subsection{5G-MEC and the Edge Storage Infrastructure (ESI)}
\label{sec_esi}

\begin{figure}
   \centerline{\includegraphics[width=.40\textwidth]{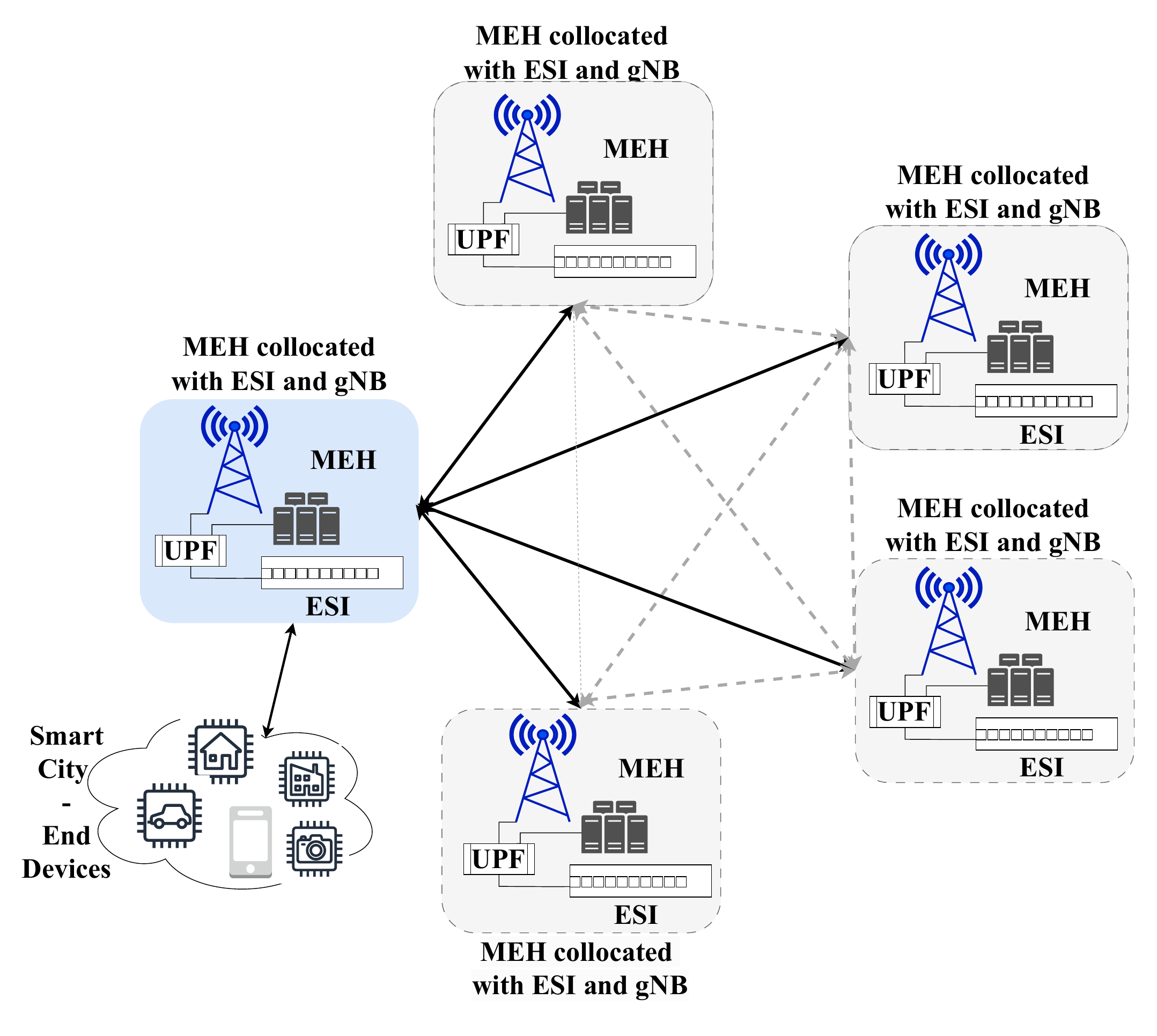}}
    \caption{Scenario overview: Smart city using MEH equipped with ESI in 5G networks}
    \label{fig:scenario}
\end{figure}

In smart cities, much of the collected data is (pre-) processed at the edge of the network~\cite{9276799}. To facilitate the overall performance in the 5G networks, a new technology called the \textit{Multi-Access Edge Computing (MEC)} moves some of the functionalities close to the end user. In the 5G-MEC setup, the MEC infrastructure deployed at the edge, known as \textit{MEC Hosts (MEHs)}, can be collocated with the 5G base stations, called \textit{gNBs}~\cite{ETSI:5G:MEC}.
A MEH is defined as an entity that provides compute, storage, and network resources for MEC applications \cite{etsi_mec003} 
When collocated with 5G base stations, the gNB can offload tasks from the MEH, e.g., by providing storage services and computing services and thus helping in lowering latency and transmission burden \cite{sun2021edge}.  

The MEHs might be further enhanced with a component to facilitate large data storage. Inspired from the literature~\cite{9077366}\footnote{The reference is not on 5G-MEC but edge in general.}, we will refer to this as the \textit{Edge Storage Infrastructure (ESI)}.
Depending on the architectural design choices, this might also be perceived as a \textit{Storage Area Network (SAN)} that is attached to the \textit{User Plane Function (UPF)} of the 5G-MEC. A SAN is a network with a special purpose - to transfer data to and from storage elements, using a communication infrastructure, and that can come in different flavors, e.g., \textit{Virtualized SAN (VSAN)}~\cite{khattar1999introduction}. Therefore, we can consider the scenario where the (pre-)processing takes place in the MEHs, but the (large) storage is outsourced to the SAN. This approach might be of particular interest in case of high storage necessities at the edge.
Note that we use the SAN as an exemplification, but any solution that allows the MEHs to directly and efficiently exchange stored data (regardless of whether the data is stored internally or externally in the MEH) is of interest. In the settings of a distributed ESI, we assume a one-to-one relation between an MEH and a storage node. We further assume that the MEH and the storage are collocated and refer to this assembly as a \textit{node}, or sometimes as simply the MEH. Thus, an MEH can access its own storage and, by case, send stored data to other MEHs. Note that although we refer to this in the settings of 5G-MEC, the solution can be suitable for MEC in general, not necessarily using the 5G connectivity.
In the following, we consider all the internals transparent, regardless of the detailed implementation.
Fig.~\ref{fig:scenario} illustrates the given settings, focusing on one single node, which is highlighted.

The usability of such an approach is of particular interest in smart cities, where devices such as sensors and cameras collect vast quantities of data, much of this data being sensitive. In the case of smart city surveillance systems, MEHs are placed between the monitoring devices and the cloud~\cite{8089336}. This way, the MEC layer only sends necessary and lesser data (e.g., summaries of data) to the cloud~\cite{8664595}.
However, by design, the MEHs are less resourceful devices and might need more data storage space to accommodate the high amount of data. Sending the data to a cloud would create latency issues and bandwidth constraints~\cite{chawla2020ai}, so using 5G-MEC to process large raw data and analyze it at the edge is preferable.
As given above, an architecture with MEHs equipped with storage functionalities such as SAN would be a way to achieve this.

\subsection{Secret Sharing}

\textit{Secret Sharing Schemes (SSS)} are cryptographic primitives that allow one party, called the \textit{dealer}\footnote{In the literature, there exist SSS without a dealer too, but we ignore those, as they are not relevant for our current purpose.} to \textit{share} a \textit{secret} into several \textit{shares} such as \textit{qualified} sets of parties can reconstruct the secret, while \textit{unqualified} set of parties cannot. We will further denote the shared secret by $s$, the space of possible secrets by $S$ (i.e., $s \in S$), the dealer by $D$, and the set of parties by $P = \{P_1, \dots, P_n\}$, with $n > 1$ the number of parties. We allow the dealer $D$ to be both one of the parties in $P$ or an external party. We formally define SSS as follows:

\begin{definition}
A Secret Sharing Scheme (SSS) $\Pi$ is a pair of algorithms $(Sh,Rec)$, where:
\begin{itemize}
    \item $Sh$ is a randomized algorithm that on input a secret $s \in S$ and a set of parties $P = \{P_1, \dots, P_n\}$, with $n > 1$ returns a set of shares $sh = (sh_1, \dots, sh_n)$, where share $sh_i$ corresponds to the party $P_i$, for all $i = 1, \dots, n$.
    \item $Rec$ is a deterministic algorithm that, on input, a subset of shares $s_A = \{sh_i|i \in A$, $A$ qualified$\} \subseteq sh$ returns $s$.
\end{itemize}
\end{definition}

\textit{Threshold SSS} is a particular implementation of SSS, for which the secret can be recovered if the number of parties participating in reconstruction is at least equal to a given threshold. Let this threshold be $k$, with $1 < k \leq n$. Then, any sub-set of participants $A \subset P$ with $card(A) \geq k$ is qualified (i.e., parties in $A$ can recover $s$), and any sub-set of participants $B \subset P$ with $card(B) > k$ is unqualified (i.e., parties in $B$ cannot recover $s$).
An SSS is \textit{perfect} if, for any unqualified set of parties $B$, it holds that $B$ cannot find any information about $s$. This means that the secret $s$ remains perfectly hidden to $B$, in the sense that the set of shares in $B$ gives no extra information about $s$.
Shamir's SSS \cite{shamir1979share} is probably the most popular threshold SSS, also providing perfect secrecy.

\begin{definition}[Shamir's SSS \cite{shamir1979share}]
Shamir's SSS $\Pi = (Sh,Rec)$ is defined as follows:
\begin{itemize}
    \item $Sh$: Choose $f \leftarrow^R Poly_{k-1}[X]$ with $f(0) = s$, where $Poly_{k-1}[X]$ is the set of polynomials of degree $k-1$ with coefficients in $S$, and $\leftarrow^R$ is the uniformly random sampling of $f$ in $Poly_{k-1}[X]$, then compute $sh_i = f(i)$, for all $i = 1, \dots, n$.
    \item $Rec$: Reconstruct $s = \sum_{i=1}^{k} sh_i \prod_{j=1,j\neq i}^{k} \frac{x_j}{x_j - x_i}$, where $A = \{sh_i, i = 1, \dots, k\}$ is qualified.
\end{itemize}
\end{definition}

Note that the polynomial $f$ is randomly chosen for each sharing, such as its degree is $k-1$ and its free term is $s$ (i.e., $f(0) = s$), the secret to being shared. Also note that, without losing generality, we stated the definition for $A$ with $card(A) = k$, the smallest set length to allow reconstruction (
more than $k$ shares will also allow reconstruction, e.g., by simply ignoring some of the shares).

\section{The Solution}
\label{sec_solution}

We use a $(k,n)$-\textit{threshold SSS} (in particular, Shamir's SSS) to build a storage system for 5G-MEC under the assumptions given in Section \ref{sec_esi}, which is simultaneously secure and fault-tolerant by construction. The data is split into $n$ shares, each being stored on a different node. The original data can only be reconstructed by combining at least $k$ (out of the $n$) shares. By construction, this approach assures functionality for up to $n-k$ inaccessible nodes. The inaccessibility of the nodes can be caused by either intentional attacks or unintentional failures.
As in \cite{al2018collaborative}, data is - to some extent - replicated at the edge, in the sense that the dimension of the overall stored secrets is increased in comparison with the initial data. However, if \cite{al2018collaborative} replicates \textit{highly accessed} data, we are now interested in sharing \textit{highly sensitive data}.

\begin{figure}
    \centerline{\includegraphics[width=.5\textwidth]{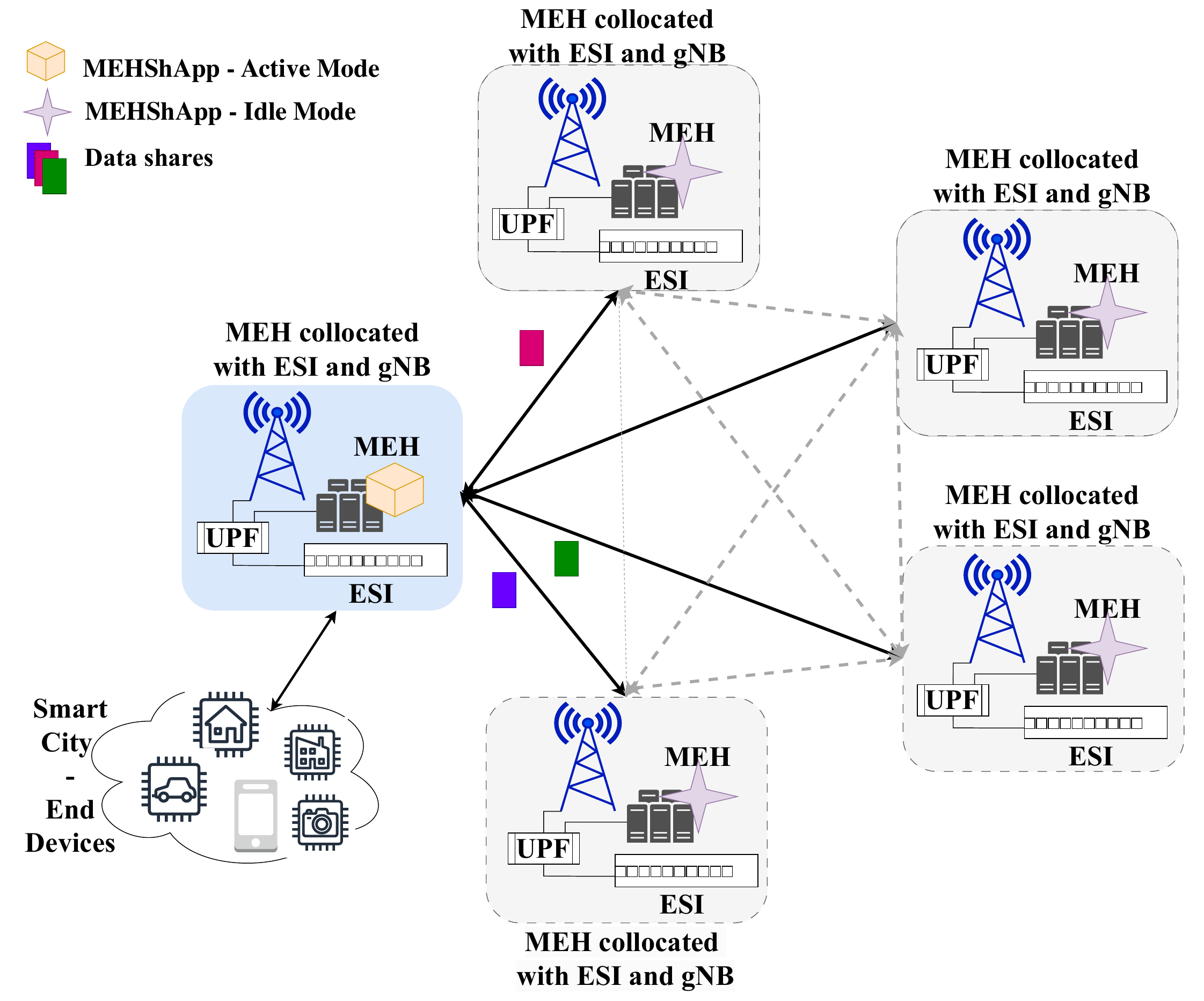}}
    \caption{\textsf{MECShApp} - Smart City scenario 
    }
    \label{fig:case}
\end{figure}

We assume that the secret sharing and reconstruction functionalities are implemented into a MEC application, further referred to as \textsf{MECShApp}. 
This application is placed on the MEHs and runs whenever triggered, to store or retrieve sensitive data originating from a User Equipment (UE) via the 5G radio interface (we refer to the UE regardless of whether it is a user device or a sensor, a camera, etc.). 
Multiple MEC applications can run simultaneously on an MEH, and continuously running \textsf{MECShApp} on all nodes becomes resource-consuming. Therefore, the \textsf{MECShApp} could accept two modes: Active and Idle, with Idle being the resource-saving mode.
Fig.\ref{fig:case} illustrates these settings.

\subsection{Goals}
\label{sec_goals}

The two main goals of the proposed solution are \textit{data secrecy} and \textit{data availablity}. Data secrecy ensures the privacy of the data in the sense that the data should remain hidden from unauthorized parties. Data availability ensures that data are available to authorized parties at request, including data recovery in case of intentional or unintentional faults. As a secondary goal, \textit{performance} ensures low overhead in data transmission, data storage, and computation cost.

\subsection{Functionality}
\label{sec_functionality}

\begin{figure}
    \centerline{\includegraphics[width=.5\textwidth]{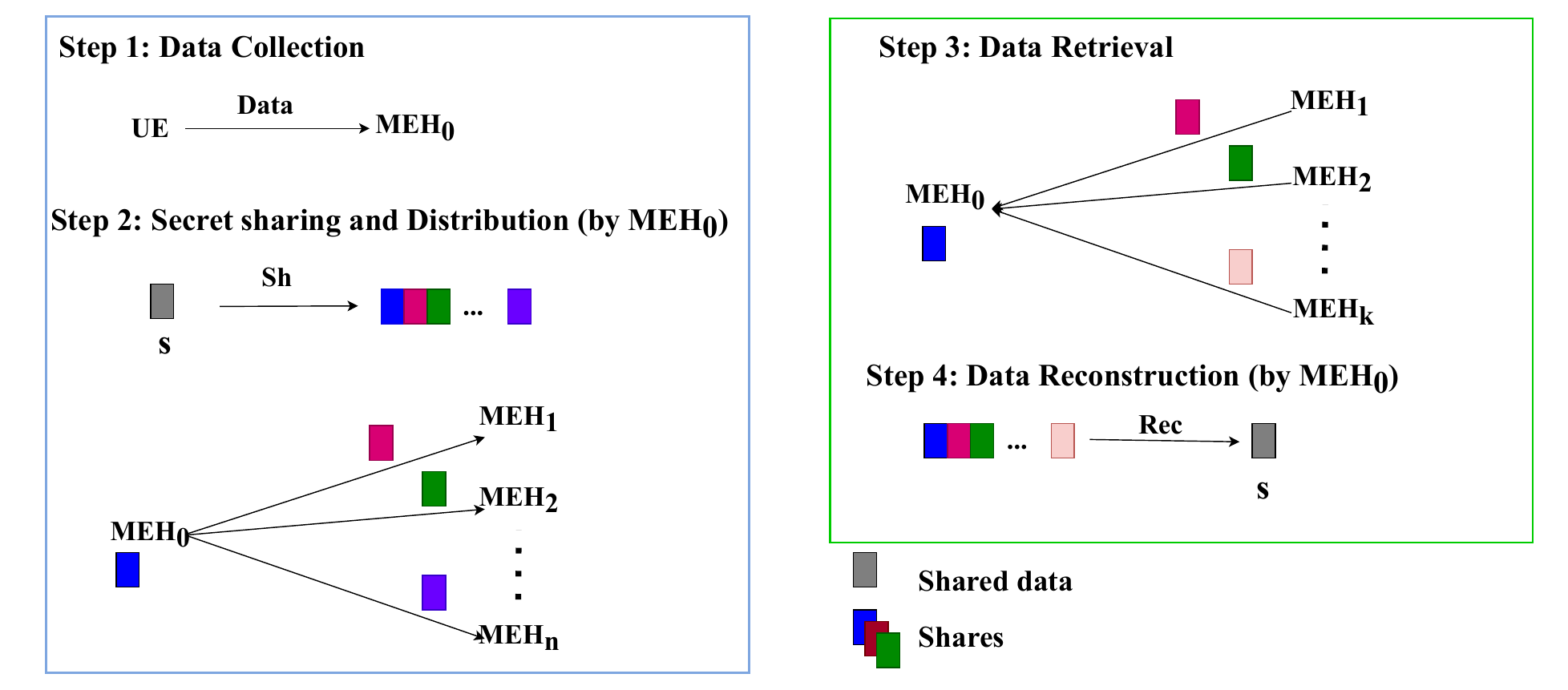}}
    \caption{Functionality of the \textsf{MECShApp}}
    \label{fig:process}
\end{figure}

Fig.~\ref{fig:process} illustrates the data sharing process performed by the \textsf{MECShApp}. The user equipment $UE$ collects the data and sends it to $MEH_0$ (step~1). Then, $MEH_0$ runs the $SSS$ algorithm to create $n$ shares. Once the shares are created, $MEH_0$ keeps one share local and sends the other $n-1$ shares to the other MEH servers (step 2).

When data retrieval is triggered, the \textsf{MECShApp} selects $k-1$ nodes $MEH_1, \dots, MEH_k$ (without losing generality, after a possible reordering) and requests their shares. In response to its request, $MEH_0$ receives the shares (step 3) and reconstructs the original data from the shares (step 4).

We have assumed that the process runs smoothly in the sense that there are no errors, interruptions, etc., and that all nodes are reachable and respond with their shares upon request.  Of course, by case,  $MEH_0$ might request more than $k$ (and up to $n$) shares from the start to avoid subsequent rounds of requests in case some of the MEHs do not answer.
We have also skipped other details, such as data encoding into a friendly representation for the SSS, as well as data division into appropriate chunks in case the dimension is over the limit and cannot be shared in one single run of the SSS. These are general to other storage solutions based on secret sharing, so one can follow the same approach as in the existing literature.

\subsection{Implementation}
\label{sec_implementation}

\begin{figure}
    \centerline{\includegraphics[width=.5\textwidth]{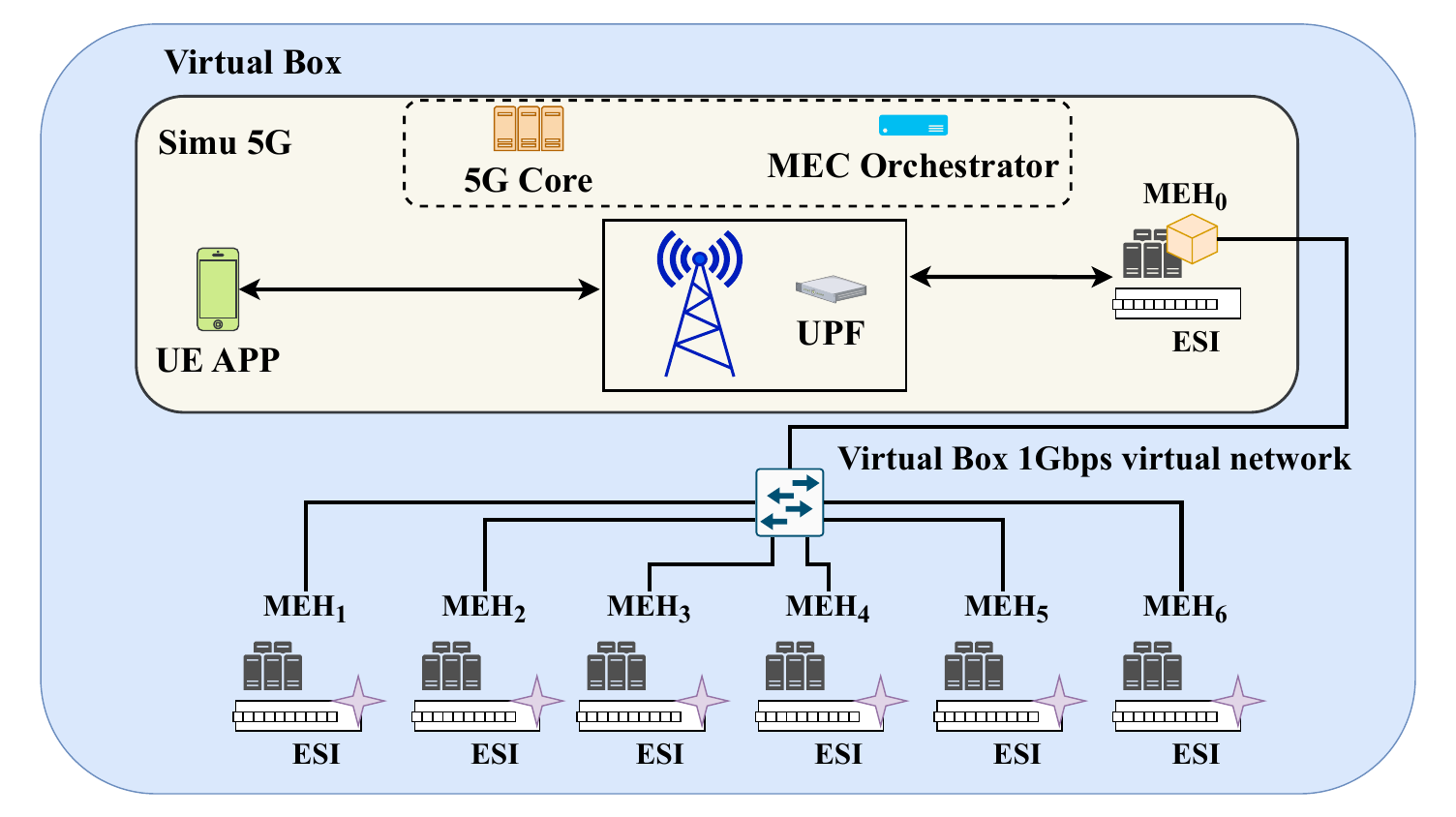}}
    \caption{Testbed setup
    }
    \label{fig:testbed}
\end{figure}

Based on the proposed functionality, we implement the testbed illustrated in Fig.~\ref{fig:testbed}. 
We use \textit{Simu5G}~\cite{nardini2020simu5g}, a well-known emulator for 5G networks, to emulate the 5G network and the MEC orchestration within the system. We further make use of a set of virtual machines acting as MEHs to emulate the processing and storage of data based on the proposed SSS-based solution. All the virtualization, including setting up Simu5G, are done within the \textit{Virtual Box} environment~\cite{virtualbox}.

When the experiment runs, a client application placed in the UE uploads some data to its connected \textsf{MECShApp}, which runs on $MEH_0$, the corresponding MEC server through the 5G network. 
Once the data is received\footnote{We ignore here if the shared data is encrypted or not before being shared, as this is transparent to the sharing itself (up to data size increase).}, \textsf{MECShApp} runs the SSS algorithm and creates the data shares. These data shares will then be sent to the other MEH servers hosting ESI service through the internal MEC network. We will further calculate the time differences between various stages of this data flow from the time the data starts to upload until the shares are stored in the network (see Section \ref{sec_eval}).

Simu5G uses several Python-based scripts to run as the UE APP and the MEC APP. We use Samba (SMB) service~\cite{carter2007using} on top of the Ubuntu operating system as the network file storage, with the underlying network being a 1Gbps virtual network within Virtual Box. We have used a Python library called \textit{pyshamir}~\cite{pyshamir} to run the secret sharing. The virtualization software, Virtual Box, was installed on an Intel Core i7 processor-powered physical computer with 32GB RAM. The Simu5G was allocated 4GB RAM, and all the other virtual machines configured as MEHs had 2GB RAM, each with Ubuntu 24.04 LTS as the operating system.

\subsection{Nodes Selection}
\label{sec_nodes_selection}

Until now, we have assumed no particular selection of the $n$ nodes to store the shares. However, in practice, $n$ is normally smaller than the total number of reachable MEHs, so we can define a selection process to identify the nodes to receive the shares. 
Trivial selections, such as the geographically closest neighbors or randomly chosen nodes, may raise security concerns or be transmission costly.
Therefore, we follow the previous work line and investigate the selection of the 5G-MEC nodes. However, unlike \cite{al2018collaborative}, we include the trustworthiness of the nodes as criteria. We define our proposal in the context of secret-shared data storage; nevertheless, the selection process can be seen as an independent and standalone process for 5G-MEC node selection in other scenarios, too. 

The process selects the best MEHs in terms of a weighted selection score $SS_{MEH_i}$ computed for each candidate MEH (generically denoted by $MEH_i$) and based on two criteria: (1) $ST_{MEH_i}$, a selection score based on the trustworthiness and (2) $SC_{MEH_i}$, a selection score based on the physical characteristics of the MEH:
\begin{equation}
    SS_{MEH_i} = w_{ST} \cdot ST_{MEH_i}(t) + w_{SC} \cdot SC_{MEH_i}(t)
\end{equation}

We work over discrete time steps, so we consider the scores functions of time. We keep the values in the score calculation weighted and assign a different weight $w$ to each term for flexibility. When changing the corresponding weight, one changes the importance given to the corresponding criteria in the score calculation.

(1) We define the MEH trustworthiness score $ST_{MEH_i}$ as a weighted sum of the inverse of $R^{att}_{MEH_i}$, the risk to successfully attack the $MEH_i$ and $ST^{sh}_{MEH_i}$, a sharing selection score used to enforce load balancing (within acceptable margins): 
\begin{equation}
    ST_{MEH_i} = w_{ST}^{att} \cdot 1/R^{att}_{MEH_i}(t) + w_{ST}^{sh} \cdot ST^{sh}_{MEH_i}(t) \\
\end{equation}
The risk of attack $R^{att}_{MEH_i}$ can be computed based on historical data and/or well-established risk procedures.
The sharing selection score $ST^{sh}_{MEH_i}$ depends on $N^{no}_{MEH_i}(t)$, the number of active shares that $MEH_i$ stores at time step $t$ and $ST^{as}_{MEH_i}(t)$, a score that depends on the access structures of the previously authorized sets and their instantiation for the active shares:
\begin{equation}
    ST^{sh}_{MEH_i}(t) = w_{ST}^{no} \cdot 1/N^{no}_{MEH_i}(t) + w_{ST}^{as} \cdot ST^{as}_{MEH_i}(t) 
\end{equation}
Thus, in the computation of $ST^{sh}_{MEH}$, we are interested in how many active shares the $MEH_i$ stores and which are the other MEHs with whom $MEH_i$ creates authorized sets. This is important because it avoids a small\footnote{Small here is, of course, dependent on the choice of the parameters $(k,n)$.} set of MEHs being responsible for (a) storing the vast majority of shares and (b) being part of a vast majority of qualified sets, respectively. In conclusion, it avoids creating a nucleus of high risk. The idea is to spread shares across the MEHs reasonably uniformly so that a successful attack against a small set of MEHs keeps the impact relatively low and avoids reconstructing a large quantity of secrets. Naturally, one could argue that the impact of the number of shares $N^{no}_{MEH_i}(t)$ in $ST^{sh}_{MEH_i}(t)$ can be modeled inside $ST^{as}_{MEH_i}(t)$; the idea to include it as a separate term is to emphasize the importance of load balancing.
In the 5G-MEC context, the computation of $ST^{sh}_{MEH_i}(t)$ is best performed by the orchestrator, which has a complete overview of the sharing process. 

Finally, in case of a known compromised MEH, we directly assign the trustworthiness score a minimal constant value $\textsf{MIN}$, which can, in theory, equal $- \infty$:
\begin{equation}
        ST_{MEH_i}(t) = \textsf{MIN}, if~\textsf{isCorrupt}(MEH_i) = 1
\end{equation}
The function $\textsf{isCorrupt}(MEH_i)$ returns 1 iff $MEH_i$ is known to be corrupt (i.e., compromised by an adversary). 

(2) The MEH physical characteristics score $SC_{MEH_i}$ depends on attributes such as the node's power capability, communication latency, etc. Examples of how to compute such scores have already been defined and can be taken from the existing literature \cite{al2018collaborative}. Similarly to other works \cite{sarahmulti}, we explicitly cover the case for which the memory capacity of the MEH becomes fully occupied in time, and, consequently, the MEH cannot store any new shares. Similarly, we explicitly consider the case for which the MEH becomes unreachable because of non-intentional faults. 
\begin{multline}
        SC_{MEH_i}(t) = \textsf{MIN}, if~(\textsf{isAvCap}(MEH_i,sh\_size) = 0 \\or~\textsf{isReachable}(MEH_i) = 0)
\end{multline}
In the above equations, the function $\textsf{isAvCap}(MEH_i,sh\_size)$ returns 1 iff $MEH_i$ has enough storage capacity to store a share (of size $sh\_size$), and the function $\textsf{isReachable}(MEH_i)$ returns 1 iff $MEH_i$ is reachable (i.e., $MEH_i$ is up and running). Note that we decouple unintentional faults from intentional attacks, which are already encompassed in the computation of the trustworthiness score $ST_{MEH_i}$.

Finally, the selection of the $n$ MEHs depends on the scores $SS_{MEH_i}$ and can follow two possible approaches: (1) select the $n$ MEHs with the highest score (this approach follows the idea in \cite{al2018collaborative}) or (2) select $n$ random MEHs from the set of MEHs whose score is above a given, acceptable threshold (this approach follows the idea in \cite{sarahmulti}).

\section{Evaluation}
\label{sec_eval}

\subsection{Security Evaluation}

\subsubsection{Adversarial Model}
\label{sec_adv_model}

We consider the \textit{honest-but-curious} model in which all parties behave as expected but aim to disclose additional information. This means that the nodes (in particular the MEHs) follow the protocol exactly but might want to disclose sensitive data that they are unauthorized for. We also restrict the adversary from tampering with the data in transit or at rest. Otherwise, this could be, for example, reduced to a dishonest MEH that fails to follow the protocol exactly but responds with fake data.
Even if an honest-but-curious attacker does not interfere with the process, he/she might observe patterns, find intermediate results, or perform any analysis to obtain an advantage
We also permit coalitions in the sense that more than one party (in particular, two or more MEHs) can collude to satisfy their adversarial goal.  

Note that SSS with enhanced capabilities could allow for stronger adversarial models. For example, malicious MEHs could try to respond with incorrect shares, or, the MecShApp could play the role of a malicious dealer and transmit incorrect shares. A way to mitigate this is to use (publicly) Verifiable Secret Sharing (VSS).

\subsubsection{Data secrecy}
Shamir's SSS is theoretically secure in the sense that regardless of the computational power and time of the adversary, he/she cannot reconstruct from an unqualified set of shares. This reduces to perfect secrecy (in theory) when less than $k$ nodes are compromised. If $k$ is chosen large enough, the probability of maliciously taking over enough nodes remains low, and the scheme is secure. In our case, this means that no more than $k-1$ MEHs running the \textsf{MecShApp} are allowed to get compromised.

Moreover, the \textsf{MECShApp} hosted on an honest-but-curious MEH is, by construction, directly exposing data that arrives to the MEH in clear. 
Nevertheless, we assume that sensitive data does not leave the UE in clear text, so a secure channel exists between the UE and the MEC. If the UE-MEH communication is encrypted, the security of the data at the MEH becomes computational. This is because someone can, with some probability, break the encryption and disclose the data. However, suppose the MEH permanently deletes the data after the sharing. In that case, the possibility of such an attack is restricted for much less time (the time from receiving the data until sharing and then deletion, respectively, for the whole lifetime of the data in the traditional case when data is stored in encrypted - not shared - form). The same reduction of the attack window in time happens at reconstruction.
Naturally, the best solution would be to implement the sharing directly on the UE. However, the feasibility of sharing directly at the UE is lower because of the heterogeneity of the UE devices and their physical constraints (e.g., sensors in smart cities).

\subsubsection{Availability}
The proposed solution works by construction when at least $k$ nodes are functional. Not relying on a single point of trust and allowing the reconstruction in the presence of up to $n-k$ nodes put down because of intentional attacks makes the solution more resilient to DoS.

\subsubsection{Maintenance} 

Strict access control mechanisms are necessary for maintaining security so that intentionally made changes do not result from unmonitored client software or users lacking authorization. Periodic data audits should occur to search for intentional errors or node manipulation that would keep the shares intact. MEC-level monitoring of security incidents globally must ensure they will detect known attacks with high probability. Investing in such proactive methods will ensure quick addressing of any anomalies, reducing risks, and safeguarding the integrity of the storage system. 

We leave out of the current discussion other aspects, such as implementation bugs, the security of the MEC application API towards the storage system, etc., which also impact the overall security of the solution.

\subsection{Dependability Evaluation}

\subsubsection{Availability}
By making use of a threshold SSS, our solution introduces dependability by design. The dependability is mainly referred to in terms of availability and reliability of the service, as well as the capacity for fault tolerance. In large-scale systems such as the 5G-MEC, where edge nodes might be deployed in multiple, distinct geographical locations, the usage of a threshold SSS, which assures that data remains available up to a threshold number of failure nodes, minimizes the impact of node failures or network unavailability. In our case, Shamir's SSS assures that data is recoverable when up to $n-k$ nodes are faulty or unavailable. Furthermore, the distribution of shares across different nodes minimizes the probability of a unique point of failure, making the system highly reliable.

\subsubsection{Parameter selection}
\label{subsec_param_selection}

The values of $k$ and $n$ have a critical impact on security and dependability, with a reasonable trade-off being necessary for efficiency. A large value of $k$ (and hence, $n$) leads to better data protection because as more shares are required to reconstruct the secret, the more difficult the attacker has to breach into more nodes. However, this also means that the system's efficiency degrades: the scheme becomes more complex, the number of involved entities is higher, the time to process the request increases, etc. On the other hand, a small $k$ improves fast data retrieval as few shares are required for rebuilding, but in turn, it may compromise security since the data is in the hands of the adversary if he gains access to only a few nodes. 
The value of $n$ and the difference $n-k$ between the total number of nodes and the minimal number of nodes necessary for reconstruction directly impact the fault tolerance. 

Naturally, the performance analysis performed for Shamir's SSS shows that as $n$ increases, the necessary time for data splitting and recovery becomes more noticeable~\cite{7381357}. When $n$ approaches higher values (e.g., above 20 or 30), the time consumed by the SSS becomes highly noticeable in contrast to the typical communication delays experienced in the considered MEH-SAN environment. The delay may negatively impact the overall efficiency of the system, especially in areas dependent on real-time high-speed data processing where latency is a critical factor. For example, if the customary MEH-SAN setup causes delays on the order of milliseconds, the added strain of high $n$ values can cause delays surpassing operational thresholds. To maintain efficient communication and mitigate latency in the distributed storage system, it is thus recommended to keep $n$ relatively low (e.g., $n$ not exceeding 10, max.20 nodes) and not sacrifice the advantages of the SSS scheme due to significant performance degradation.

\subsubsection{Node allocation}

Normally, the number of shares is lower than the total number of nodes in the network. Thus, the shares are generally stored in a subset of nodes. 
The decision on the nodes to accommodate the shares should be based on attributes such as the workload, the reliability and the degree of trustworthiness of the node, and the geographical location. Ideally, shares should be assigned to less-loaded nodes to prevent all nodes from being clogged with data and minimize processing delays within certain nodes. Moreover, focusing on nodes with high available capabilities or good connectivity records can help avoid a data loss issue. Geographic diversity will help protect some nodes hosting shares in the event of multiple nodes failing in the same geographical area due to causes like natural disasters. Thus, the optimal approach can be to keep track of the current load and network delay of each MEH and then continuously allocate shares to nodes that have lesser load, higher availability, and geographical dispersion to improve the robustness of the SSS-based MEC storage solution.

\subsubsection{Maintenance} 

Determining the dependability of a system depends on its real-time identification of node failures through monitoring and fault tolerance functionality. The system must be able to alert administrators when a storage node fails or is suspended and to redistribute shares to keep availability up. Continuous monitoring identifies equipment, software, or network problems for quick corrective steps. All MEC operators should consider proactive monitoring for faults and failures from the MEC system level, as it will help the administrators and users have high dependability within the MEC environment.

\subsection{Performance Evaluation}

\subsubsection{Theoretical performance}

\begin{table*}[ht]
\caption{Theoretical performance analysis.}\label{tab:1} \centering
\begin{tabular}{|p{0.2\textwidth}|p{0.08\textwidth}|p{0.62\textwidth}|}
  \hline
    \textbf{Parameter} & \textbf{Complexity} & \textbf{Explanation} \\
  \hline
  Computation (Shamir's SSS) &  $O(n log^2n)$  & Shamir's SSS requires $O(n log^2n)$ time complexity caused by the polynomial interpolation \cite{shamir1979share}. \\
Total Storage Requirement & $O(n|s|)$ & Each of the $n$ shares is of the same size $|s|$ as the original data $s$, resulting in a total storage requirement of $O(n|s|)$. \\
Total Communication Overhead (Sharing) & $O(n|s|)$ & The communication overhead is $O(n|s|)$ because all $n$ shares have to be transferred and stored across the network, and each share is the size of the shared data $|s|$.\\
Total Communication Overhead (Reconstr.) & $O(k|s|)$ & The communication overhead is (under best conditions, assuming no loss) $O(k|s|)$ because $k$ shares have to be transferred across the network, and each share is the size of the shared data $|s|$. \\
Fault Tolerance & $n - k$ & The system can tolerate up to $n - k$ node failures. \\
Maximum no. of compromised nodes (shares) & $k-1$ & Shamir's SSS enforces that $k-1$ or less shares leak no information about the shared data. \\

  \hline
\end{tabular}
\end{table*}

Table~\ref{tab:1} lists the complexities of the proposed solution and the underlying Shamir SSS~\cite{7381357,shamir1979share} when it is applied to 5G and MEC-ESI environment, particularly relative to distributed storage and secure data management. It is evident that while the security and dependability of this scheme are noteworthy, the computational resources needed for the sharing escalate as the number $n$ of the shares rises. It is imperative to determine values for $n$ (and $k$) accurately based on the specifications of an application in terms of latency and performance. The $O(n|s|)$ linear storage need exemplifies the intrinsic redundancy of the scheme, which, while also ensures fault tolerance (for up to $n - k$ nodes), it also elevates storage demands. In much the same way, the time taken for communication $O(n|s|)$ and $O(k|s|)$ requires that, in networks with restrictions on bandwidth or increased latency, diligent planning is needed to steer clear of performance limitations. The underlying SSS scheme ensures that $k-1$ or less shares will not leak information, making it a good choice for times when confidentiality of data is important, providing support for use in applications that require high levels of data protection across the 5G-MEC environment.

\subsubsection{Experimental performance}

To evaluate the performance, we look into different timings necessary to process, distribute, and store the data. Therefore, the time values listed in Table~\ref{tab:time} consider the case with and without the SSS. All values are in seconds, and their definition is the following:
\begin{enumerate}
    \item \textbf{$T_{5G}$}: time for the data to travel through the 5G network
    \item \textbf{$T_{SSS+LS}$}: time to run the SSS sharing algorithm and to store the shares on the local storage 
    \item \textbf{$T_{SSS+NS}$}: time to run the SSS sharing algorithm and to store the shares on the network, i.e., on the ESI nodes
    \item \textbf{$T_{LS}$}: time to store the data locally without running the SSS
\end{enumerate}

We chose the parameters $n=5$ and $k=3$ for the evaluation because this combination reflects a proper balance between redundancy and security. Having $n=5$, the system can tolerate up to $n-k=2$ node failures, ensuring that the system remains functional even if a few shares are lost. The threshold of $k=3$ ensures that fewer shares cannot reconstruct the secret, protecting it against attackers that compromise up to two nodes. We avoided extreme cases such as large $n$, which might introduce excessive computing resources, and very small $n$, which directly reduces the redundancy, increasing the risk of data loss due to node failures. Refer to Section \ref{subsec_param_selection} for a more detailed discussion on parameter selection.

The implementation of \textsf{MECshApp} was performed on top of Simu5G's default MEC APP template. This introduces a limitation on how much data can be sent at a time, thus limiting our tests to a maximum of 64 bytes. Testing for more extensive data is a possible subject of further work.

\begin{table}[b!]
\caption{Theoretical performance analysis.}\label{tab:time} \centering
\begin{tabular}{|l|c|c|c|c|}
  \hline
    & \textbf{25 Bytes} & \textbf{32 Bytes} & \textbf{40 Bytes} & \textbf{64 Bytes}\\
  \hline
\textbf{$T_{5G}$} [s] & 0.010689 & 0.011251 & 0.012151 & 0.012949 \\
\textbf{$T_{SSS+LS}$} [s] & 0.003565 & 0.003751 & 0.004352 & 0.004727 \\
\textbf{$T_{SSS+NS}$} [s] & 0.097281 & 0.111844 & 0.136104 & 0.148803 \\
\textbf{$T_{LS}$} [s] & 0.000305 & 0.000408 & 0.000781 & 0.000895 \\

  \hline
\end{tabular}
\end{table}

Table~\ref{tab:time} indicates that the time to perform the sharing and to store the shares locally $T_{SSS+LS}$ is significantly more time-consuming than the local storage operation $T_{LS}$. This is a natural result (also because $T_{SSS+LS}$ includes the necessary time to store the $n=5$ shares locally, so 5x more time for storing only). 
However, even if the introduction of SSS is substantial, increasing $T_{SSS+LS}$ with one order of magnitude over $T_{LS}$, in our experimental settings, we observe that secret sharing itself seems less time-consuming than the amount of time required to transmit the shares over the network and store them to the corresponding nodes. The motivation here is twofold: (1) we have only shared low-length data (up to 64 bytes), and (2) the communication between the storage nodes is done within the virtual network. We expect secret sharing to become high-cost for large data and communication costs to be lower in real networks. Nevertheless, the experiment shows that the overall delay is acceptable, which makes the solution viable for usage in scenarios like the ones mentioned.

\section{\uppercase{Conclusions}}
\label{sec_conclusion}
We have investigated a method to improve the secure data storage in edge nodes associated with 5G-MEC to provide data secrecy, high resilience to failures, and elevated data availability. With the focus on protecting sensitive data, our scheme uses Shamir's scheme to implement this secure storage method as an MEC application. Although the approach is not novel in terms of ideas (secret sharing has been used to secure storage before), it brings novelty in several directions. We have reviewed the applicability of secret sharing in 5G-MEC, including the discussion of existing storage solutions, performed an experimental evaluation, and proposed a node selection mechanism to decide on the nodes to store the shares while considering the trustworthiness of the MEHs as a main criterion. 
As future contributions, more experimental results should be obtained to assess such an MEC application's usability by measuring the performance across different operational conditions. 

\bibliographystyle{IEEEtran}
\bibliography{listofpubs}

\end{document}